\documentclass[3p]{elsarticle}

\usepackage{lineno,hyperref}
\usepackage{xspace}
\modulolinenumbers[1]
\usepackage{bm}
\usepackage{amssymb}
\usepackage{hyperref}
\usepackage{amsmath}
\usepackage{graphicx}
\usepackage{comment}
\usepackage{color}

\newcommand{\s}{\ensuremath{\sqrt{s}}\xspace}
\newcommand{\spher}{\ensuremath{S_{\rm{0}}}\xspace}
\newcommand{\py}{PYTHIA~8\xspace}
\newcommand{\ep}{EPOS~LHC\xspace}
\newcommand{\pt}{\ensuremath{p_{\rm{T}}}\xspace}
\newcommand{\meanpt}{\ensuremath{\langle p_{\mathrm{T}} \rangle}\xspace}
\newcommand{\mult}{\ensuremath{N_{\mathrm{ch}}}\xspace}
\newcommand{\gevc}{\ensuremath{\rm{GeV/}c}\xspace}

\journal{Journal of \LaTeX\ Templates}

\begin{document}

\begin{frontmatter}

\title{{\bf Understanding the transverse-spherocity biased data from pp collisions at the LHC energies}}

\author{Antonio Ortiz}
\address{Instituto de Ciencias Nucleares, Universidad Nacional Aut\'onoma de M\'exico, \\ Apartado Postal 70-543,
Ciudad de M\'exico 04510, M\'exico}

\author{Lizardo Valencia Palomo and Victor Manuel Minjares Neriz}
\address{Departamento de Investigaci\'on en F\'isica, Universidad de Sonora, \\ Blvd. Luis Encinas y Rosales S/N, Col. Centro, Hermosillo, Sonora, M\'exico}


\begin{abstract}

 The ALICE collaboration recently reported the mean transverse momentum as a function of charged-particle multiplicity for different pp-collisions classes defined based on the ``jettiness'' of the event. The event ``jettiness'' is quantified  using transverse spherocity that is measured at midpseudorapidity ($|\eta|<0.8$) considering charged particles with transverse momentum within $0.15<p_{\rm T}<10$\,GeV/$c$. Comparisons to \py (tune Monash) predictions show a notable disagreement between the event generator and data for jetty events that increases as a function of charged-particle multiplicity. This paper reports on the origin of such a disagreement using \py event generator. Since at intermediate and high \pt ($2<\pt<10$\,GeV/$c$), the spectral shape is expected to be modified by color reconnection or jets, their effects on the average \pt are studied. The results indicate that the origin of the discrepancy is the overpredicted multijet yield by \py which increases with the charged particle multiplicity. This finding is important to understand the way transverse spherocity and multiplicity  bias the pp collisions, and how well models like \py reproduce those biases. The studies are pertinent since transverse spherocity is currently used as an event classifier by experiments at the LHC. 
 
\end{abstract}


\end{frontmatter}


\section{Introduction}

Quantum Chromodynamics (QCD) predicts that at very high-energy densities, ordinary nuclear matter undergoes a crossover transition to primordial hot QCD matter (deconfined quarks and gluons). This state of matter, called quark--gluon plasma (QGP), existed a few microseconds after the Big Bang~\cite{QGP2018,ALICEjourney}. The QGP can be recreated in the laboratory through high-energy heavy-ion collisions~\cite{PHENIX:2004vcz,STAR:2005gfr}. For this purpose, heavy ions have been collided at ultra-relativistic center-of-mass energies per nucleon pair at RHIC and LHC. LHC data support the formation of a medium with a lower-bound energy density between 10 and 20\,GeV/(fm$^2 c$) and effective temperature of almost 300\,MeV~\cite{ALICEenergy,ALICEtemperature}. This fireball has shown to be a strongly-interacting system of quarks and gluons with very low viscosity-to-entropy ratio (nearly an ideal fluid) that presents a hydrodynamic behaviour~\cite{ALICEflow1,ALICEflow2}. Once created, the system expands and cools down very fast with a characteristic decoupling time of approximately 10\,fm/$c$~\cite{ALICEproptime}. Among the observables used to study the QGP, event-by-event fluctuations like the number of charged particles (multiplicity) or the mean transverse momentum ($\langle p_{\mathrm{T}} \rangle$) are important to understand the dynamical variations associated to the formation of the medium~\cite{ALICE:2014gvd}. The collective expansion of the system is responsible for the shape of the \pt spectra at low- and intermediate-transverse momentum ($p_{\mathrm{T}} \lesssim4$\,GeV/$c$)~\cite{ALICEptPbPb}. Above this threshold, the distributions are a consequence of the initial hard scatterings in the collisions. On top of this, jet quenching originated by the energy loss of partons traversing the dense medium created by the collision, remains as a key observable in the study of QGP~\cite{JetQuenching}.

Before the start of the LHC, proton-proton (pp) and proton-lead collisions (p-Pb), the so-called small collision systems, were simply treated as control experiments. However, one of the most unexpected results in the high-energy physics area in the last 15 year has been the discovery of QGP-like effects in high-multiplicity pp and p-Pb interactions. It all started with the observation of a long-range, near-side dihadron correlation (ridge structure) in pp collisions and afterwards also discovered in p-Pb collisions~\cite{Ridgepp,RidgepPb}. Since then, more and more collective effects in small systems have been unveiled~\cite{SmallSystemsReview}. The physics beneath these observations have been part of an intense debate. Several theoretical approaches have been suggested to explain the QGP-like effects in small-collision systems. For example, in the range of applicability the color-glass condensate effective field theory, the flow-like behavior can be produced in the early stages of the collision~\cite{Bzdak:2013zma,Dusling:2012cg}. Alternatively, the effects could develop during the collective evolution, where hydrodynamics is applicable~\cite{Mantysaari:2017cni,Greif:2017bnr,HydroModel}. Other approaches implemented in Monte Carlo event generators like \py employ effects such as color reconnection or rope hadronization to perform a microscopic description of the system~\cite{SaturationDihadrons,OverlappingStringsModel}. Indeed, it has been shown that such models and implementations in \py can produce flow-like effects in pp collisions~\cite{FlowlikeEffects}. 

Charged particle multiplicities ($N_{\mathrm{ch}}$) and $p_{\mathrm{T}}$ distributions have been extensively studied by experiments at the LHC and at RHIC in small systems \cite{STAR:2006axp,PHENIX:2011rvu,ALICE900GeV,CMS0.9To7TeV,ATLASmultiplicities,ALICE13TeV,ALICE:2023csm}. The data indicate a clear increase of the $\langle p_{\mathrm{T}} \rangle$ with increasing multiplicity. On one hand, this phenomenon is described in \py by multiparton interactions (MPI) and allowing the interaction among partons before the hadronization via color strings (color reconnection), thus hardening the $p_{\mathrm{T}}$ spectra at intermediate $p_{\rm T}$ but decreasing the average multiplicity~\cite{PythiaBeyondLeadingColour}. On the other hand, in \ep, an event generator featuring core-corona effect, the increase of the \meanpt is determined by the collective expansion of the system \cite{EPOSLHC}.

One issue when pp collisions are analyzed as a function of the charged particle multiplicity (measured in a narrow pseudorapidity interval) is that high-multiplicity events are biased towards multijet final states. The effect is illustrated when the $p_{\rm T}$ spectra of high-multiplicity pp collisions is normalized to the analogous quantity measured in minimum-bias pp collisions. The ratio shows a continuous rise with increasing $p_{\rm T}$ and it gets steeper for larger multiplicity values~\cite{ALICE:2019dfi}. One way to mitigate the bias was proposed some years ago, the idea consists in measuring the ``jettiness'' of the event using event shape observables like transverse spherocity~\cite{Ortiz:2015ttf,Ortiz:2017jho,Ortiz:2022mfv}. Since then, different measurements have been reported using event shape selections~\cite{ALICE:2023bga,ALICE:2023zbh}. The ALICE collaboration reported the first measurement of the \meanpt as a function of multiplicity for different spherocity classes~\cite{ALICE:2019dfi}. While results for minimum-bias and high-spherocity pp collisions (isotropic events) were well described by \py tune Monash, the agreement was broken for low-spherocity pp collisions (jetty events). This was a surprise since Monash was obtained from a tune to LHC data, and therefore is known to describe several observables of unidentified charged particles~\cite{MonashTune}. In this paper, the origin of this discrepancy is studied focusing on color reconnection as it is known to modify the $p_{\rm T}$-spectral shape  at intermediate \pt ($2<\pt<4$\,GeV/$c$). Since in this \pt range, the transition between soft and hard processes occurs, the impact of color reconnection and jets is also explored. Another motivation is the bias of the high-multiplicity pp sample towards multijet final states. 

This paper is organized as follows: section 2 provides a brief description of event shapes and Monte Carlo event generator. Section 3 explores the origin of the difference between data and PYTHIA 8, as well as the impact of jets to reconcile the event generator and data. Finally section 4 summarizes the conclusions.

\section{Spherocity and PYTHIA~8 event generators}

Event shapes are sensitive to the spacial distribution of particles produced in a collision. They have been extensively used to characterize QCD in electron-positron collisions, as for example in the extraction of the strong-coupling constant, understanding hadronization process or even in the parton shower tuning in event generators~\cite{EventShapes}. In hadronic interactions, event shapes are restricted to the transverse plane relative to the beam direction, making the observables insensitive to the boost along the beam direction~\cite{OrtizVelasquez:2009pey}.

Among the event shapes currently in use, transverse spherocity (\spher) has shown to be a good tool to classify the high-multiplicity pp collisions as either multijet final states (jetty) or  uniform particle emission (isotropic)~\cite{ALICE:2023bga}. Jetty events are associated to hard partonic scaterings while isotropic events are related to collisions in which several semi-hard parton-parton scatterings occurs within the same pp collision. Transverse spherocity, from now on simply called spherocity, is defined relative to a unit vector ($\hat{n}$) that minimizes the ratio:

\begin{equation}
S_0 = \frac{\pi^2}{4} \left(  \frac{ \sum_i \mid \vec{p}_{\mathrm{Ti}} \times \hat{n} \mid}{\sum_i p_{\mathrm{Ti}}} \right)^2.
\end{equation}
Spherocity is a normalized quantity, and as a consequence has extreme values 0 and 1, corresponding to jetty and isotropic events, respectively. Since spherocity is implicitly multiplicity dependent, in order to disentangle the multiplicity from the event shape effects, the analysis has to be double differential.  For a fixed multiplicity value, the multiplicity effect gets factorized and therefore any modification on particle production can be attributed to the event-shape selection.  Nonetheless, selecting pp collisions with high charged-particle multiplicity biases the sample affecting different observables. For instance, the neutral-to-charged kaon ratio is known to decrease when the midrapidity charged-particle multiplicity increases~\cite{ALICE:2018pal}. In \py, events with isotropic distribution of particles are associated to high underlying-event (UE) activity and therefore with a large number of MPI. Contrarily, the UE activity decreases when  spherocity is reduced.

As different physics aspects can not be extracted from theoretical grounds, event generators rely on some sets of tuned parameters to enhance the predictive power. In this paper the tune Monash, a set of parameters extracted from ${\rm e^+ e^-}$ and ${\rm p\bar{p}}$ data, is used in the \py simulations~\cite{Pythia8,MonashTune}.  During the developing parton shower of a hadronic collision, quarks and gluons are connected by colored strings. Originally, string models were based on the leading color approximation where generated partons were colored connected only to their parent emitters. In this sense, the products from different MPIs were kept independent. Color reconnection (CR) allows the interaction among partons from originally non-correlated MPIs, implying a far richer colored topology than the original leading color method~\cite{EventGenerators}.  The MPI-based color reconnection model implemented in \py Monash sets the probability of a low-\pt parton to be merged with a higher-\pt one, in such a way that the total string length gets minimized. The CR mechanism is governed by a parameter called reconnection range ($RR$), with a value of 1.8 for the Monash tune. 

\section{Average transverse momentum as a function of multiplicity and spherocity}

In order to compare with available ALICE data, only pp collisions producing at least one charged particle with $p_{\rm T}>0$ in the pseudorapidity interval $|\eta|<1$ are used, this particular selection is called INEL$>$0. Furthermore, particles are required to be promptly produced in the collision, including all decay products excluding those from weak decays. Following the experimental definition of spherocity, the minimum number of particles in the selected event must be at least three and a restricted \pt range of 0.15 to 10\,GeV/\textit{c} is demanded~\cite{ALICE:2019dfi}. With this information it is possible to compute a \spher distribution for each event multiplicity and derive the \spher percentiles corresponding to each \mult interval. Events falling within the 0-10$\%$ spherocity class are labeled as jetty, while those in the opposite extreme range of 90-100$\%$ are the istropic events. Finally, the \meanpt is computed per event multiplicity in each \spher class. 

\begin{figure*}
\begin{center}
\includegraphics[width=0.49\textwidth]{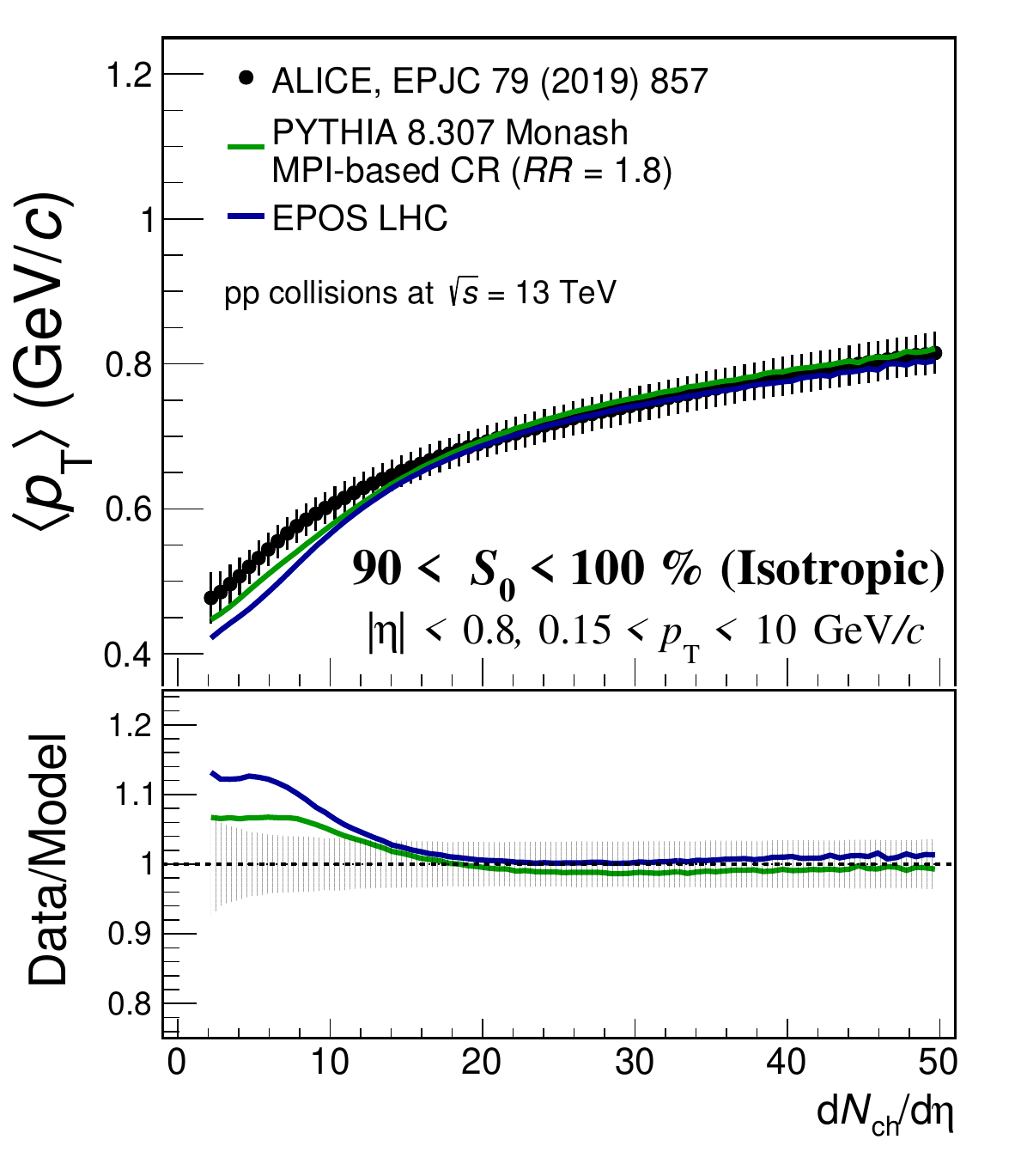}
\hspace{-0.4cm}
\includegraphics[width=0.49\textwidth]{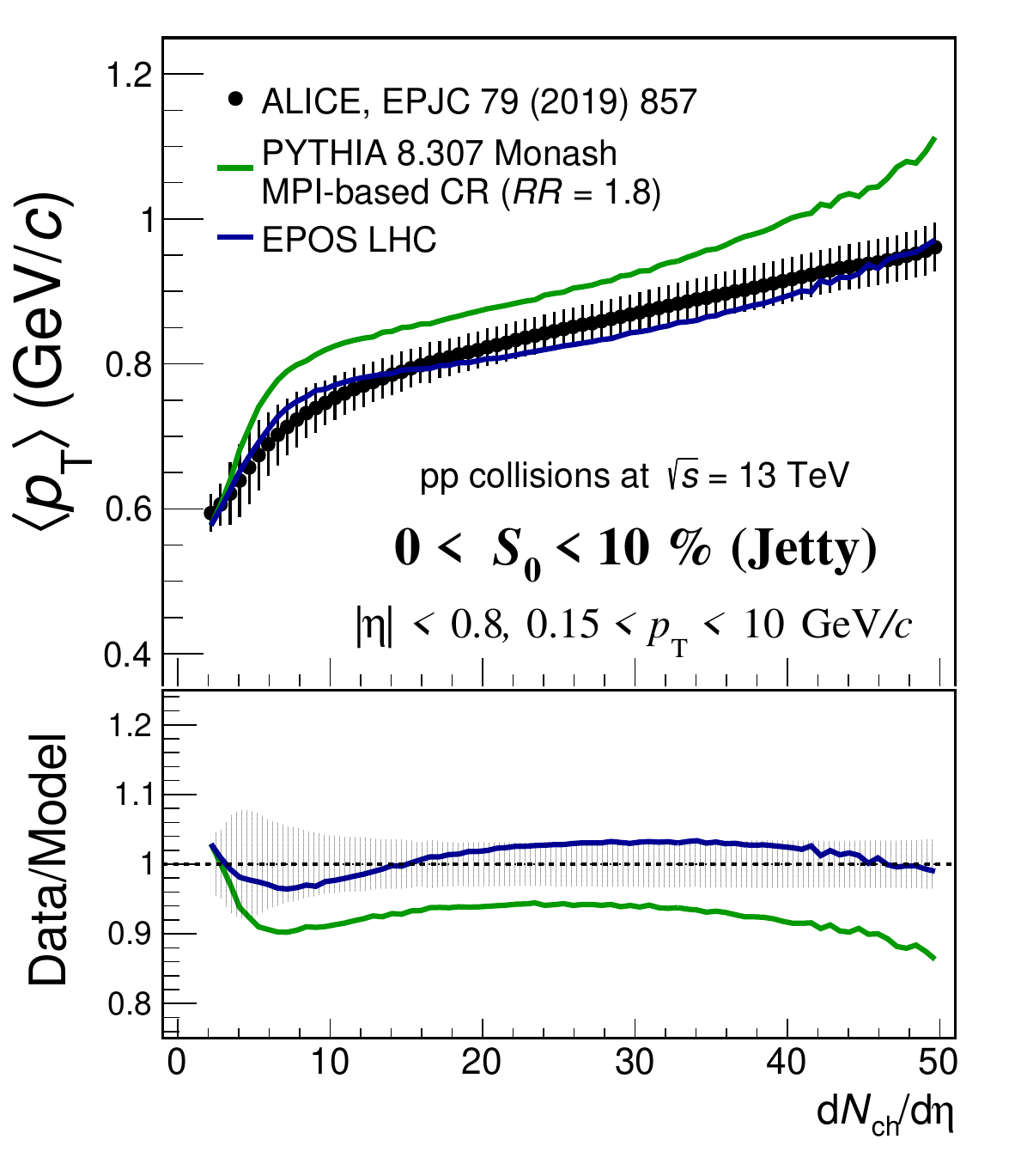}
\caption{Average transverse momentum as a function of d$N_{\mathrm{ch}}$/d$\eta$ for pp collisions at \s = 13 TeV for two different spherocity classes: isotropic (\textit{left}) and jetty (\textit{right}). Black points correspond to data and error bars are the associated systematic uncertainties. Data are compared to two different Monte Carlo predictions (solid lines). Green line is the result from \py Monash, while blue line is the result from \ep. Bottom panel presents the data-to-model ratio, where the shaded area around unity is the systematic uncertainty.}
\label{OriginalModels}
\end{center}
\end{figure*}

\subsection{Color reconnection effects}

Figure \ref{OriginalModels} presents the \meanpt as a function of d$N_{\mathrm{ch}}$/d$\eta$ for isotropic and jetty events. Data are compared with \py Monash and \ep predictions~\cite{EPOSLHC}, the results are fully consistent with those reported in Ref.~\cite{ALICE:2019dfi}. For isotropic events both event generators can correctly reproduce the data when d$N_{\mathrm{ch}}$/d$\eta$ $>$ 12. For jetty events only \ep can describe the data within the full multiplicity range. Indeed, \py Monash completely falls away from the systematic uncertainties. Furthermore, the data-to-model ratio shows a minimum at d$N_{\mathrm{ch}}$/d$\eta$ $\approx$ 10 followed by a maximum and then a fast increasing divergence. It is worth mentioning that \ep is based on the Parton-Based Gribov-Regge Theory, where nucleon-nucleon collisions are addressed at the parton level via Pomeron exchanges that generate a parton ladder~\cite{EPOSLHC}. These are in turn considered as flux tubes or strings that can decay via the emission of quark-antiquark pairs. \ep implements a core-corona effect, where core region presents a larger density of strings relative to the corona. The event generator is tuned to reproduce the collective effects observed in the small systems at the LHC.

\begin{figure*}
\begin{center}
\includegraphics[width=0.49\textwidth]{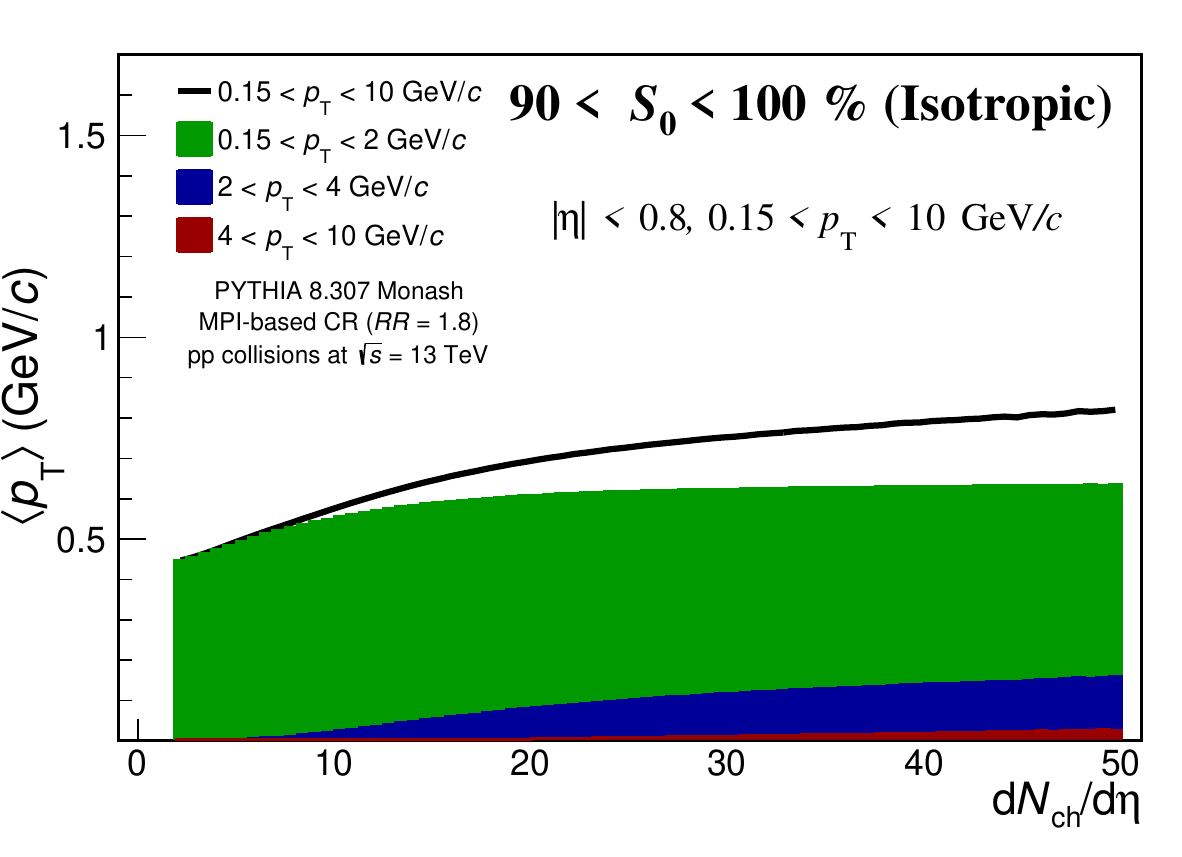}
\hspace{-0.4cm}
\includegraphics[width=0.49\textwidth]{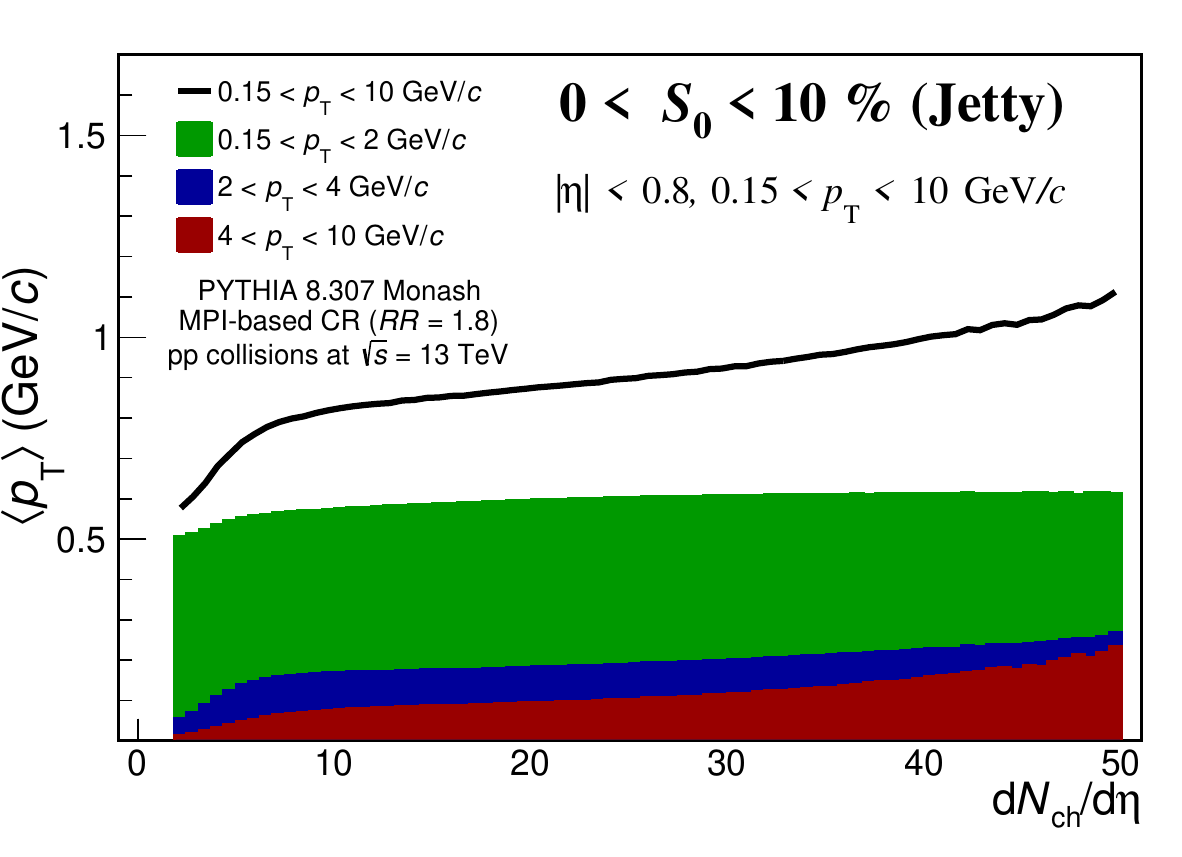}
\caption{Average \pt as a function of d$N_{\mathrm{ch}}$/d$\eta$ for pp collisions at \s = 13 TeV for two different spherocity classes: isotropic (\textit{left}) and jetty (\textit{right}). Black line is the \pt-integrated distribution, green area is for low-\pt particles with $0.15<p_{\rm T}<2$\,GeV/$c$, blue for intermediate-\pt particles with $2<p_{\rm T}<4$\,GeV/$c$, and red area for high-\pt particles with $4<p_{\rm T}<10$\,GeV/$c$.}
\label{ParticleComposition}
\end{center}
\end{figure*}

In order to understand the disagreement between \py Monash and data for jetty events, the contribution of particles within different \pt intervals to \meanpt is studied. Figure~\ref{ParticleComposition} presents the \meanpt as a function of d$N_{\mathrm{ch}}$/d$\eta$ for isotropic and jetty events. The contribution of particles with transverse momentum within $0.15< p_{\rm T} < 2$\,GeV/\textit{c}  (low-\pt particles),  $2 < p_{\rm T} < 4$\,GeV/\textit{c} (intermediate-\pt particles) and $4 < \pt < 10$\,GeV/\textit{c} (high-\pt particles) is shown. Isotropic events are fully dominated by low-\pt particles up to d$N_{\mathrm{ch}}$/d$\eta$ $\approx$ 30. As a result, below this threshold the shape of \meanpt closely resembles the shape dictated by the most inclusive \meanpt. For high-multiplicity events the increase of the \meanpt is influenced by intermediate-\pt particles. Jetty events show a completely different behavior. Up to d$N_{\mathrm{ch}}$/d$\eta$ $\approx$ 30, the shape of the most inclusive \meanpt is mostly due to low- and intermediate-\pt particles, but there is also a non-negligible contribution from high-\pt particles. Finally, the fast increase of the \meanpt in the high-multiplicity regime (the third rise of the average \pt as a function of multiplicity) is mainly driven by high-\pt particles.

\begin{figure*}
\begin{center}
\includegraphics[width=0.49\textwidth]{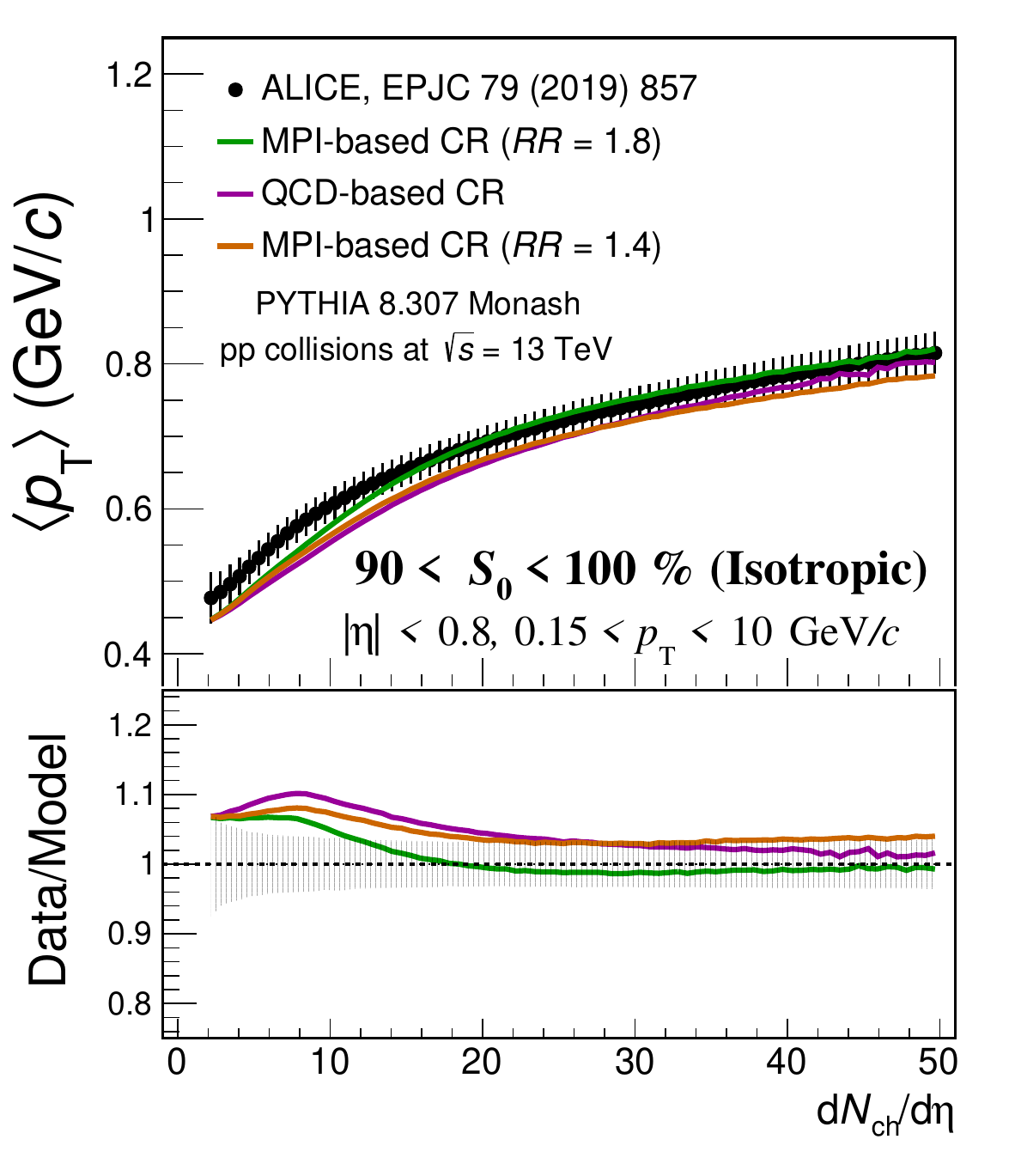}
\hspace{-0.4cm}
\includegraphics[width=0.49\textwidth]{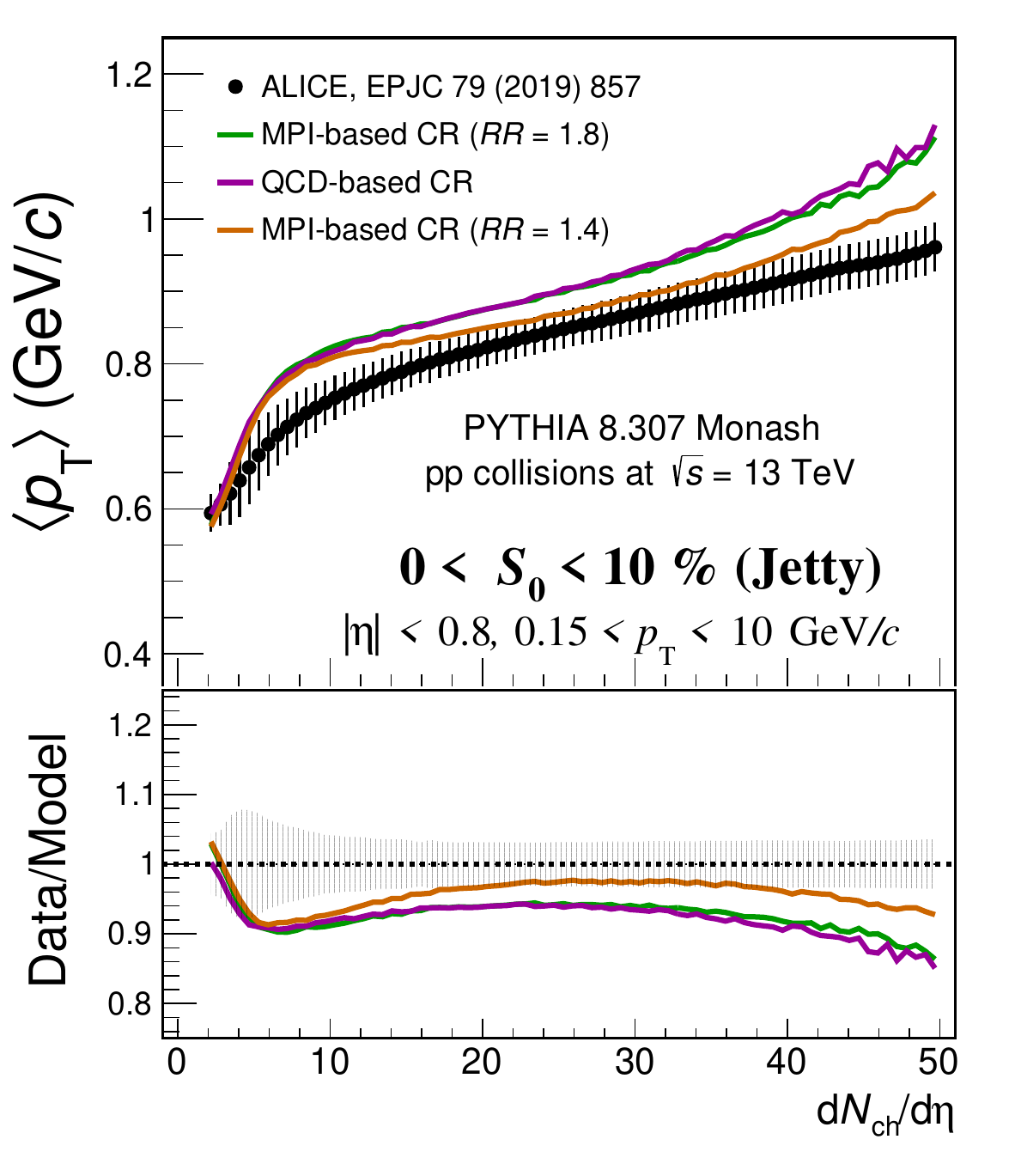}
\caption{Average \pt as a function of d$N_{\mathrm{ch}}$/d$\eta$ for pp collisions at \s = 13 TeV for two different spherocity classes: isotropic (\textit{left}) and jetty (\textit{right}). Black markers correspond to data and error bars are the associated systematic uncertainties. Data are compared to three different Monte Carlo predictions (solid lines). Green line is the result from \py Monash (default $RR=1.8$), purple represents the \py color reconnection based on a QCD model, and orange line is \py Monash ($RR = 1.4$). Bottom panel presents the data to model ratio, where the shaded area around unity is the systematic uncertainty.}
\label{NewModels}
\end{center}
\end{figure*}

Since CR is known to modify the \pt spectral shape, different CR models were tested. The first one employs a new CR method available in \py that is based on QCD rules to determine the string length minimization. This model allows the formation of topological structures (junctions) when three colored strings meet at a single point. This implies that baryon production is enhanced with respect to the default CR approach~\cite{NewCR}. The simulation with this method is labeled as \py Monash (QCD-based CR) in Fig.~\ref{NewModels}. The agreement with data in isotropic events gets worst in particular for d$N_{\mathrm{ch}}$/d$\eta$ $<$ 25. For jetty events there is basically no difference relative to \py Monash. The second simulation is also based on \py Monash but with a reconnection range value of 1.4, slightly smaller than the default one. This is labeled as \py Monash ($RR = 1.4$) in Fig.~\ref{NewModels}. Lowering the $RR$ reduces the reconnection probability, increasing the particle multiplicity and decreasing the \pt of the emitted particles. As a consequence, the average transverse momentum of the event is also reduced. Results confirm this statement both for isotropic and jetty events, \meanpt from \py Monash ($RR = 1.4$) is systematically below the prediction from \py Monash ($RR=1.8$). This slight modification in the $RR$ greatly improves the agreement with data in jetty events, however, the third rise of \meanpt is till observed. For isotropic events the \meanpt predicted by \py Monash ($RR = 1.4$) is worst in the full multiplicity interval relative to the default case.

\begin{figure*}
\begin{center}
\includegraphics[width=0.5\textwidth]{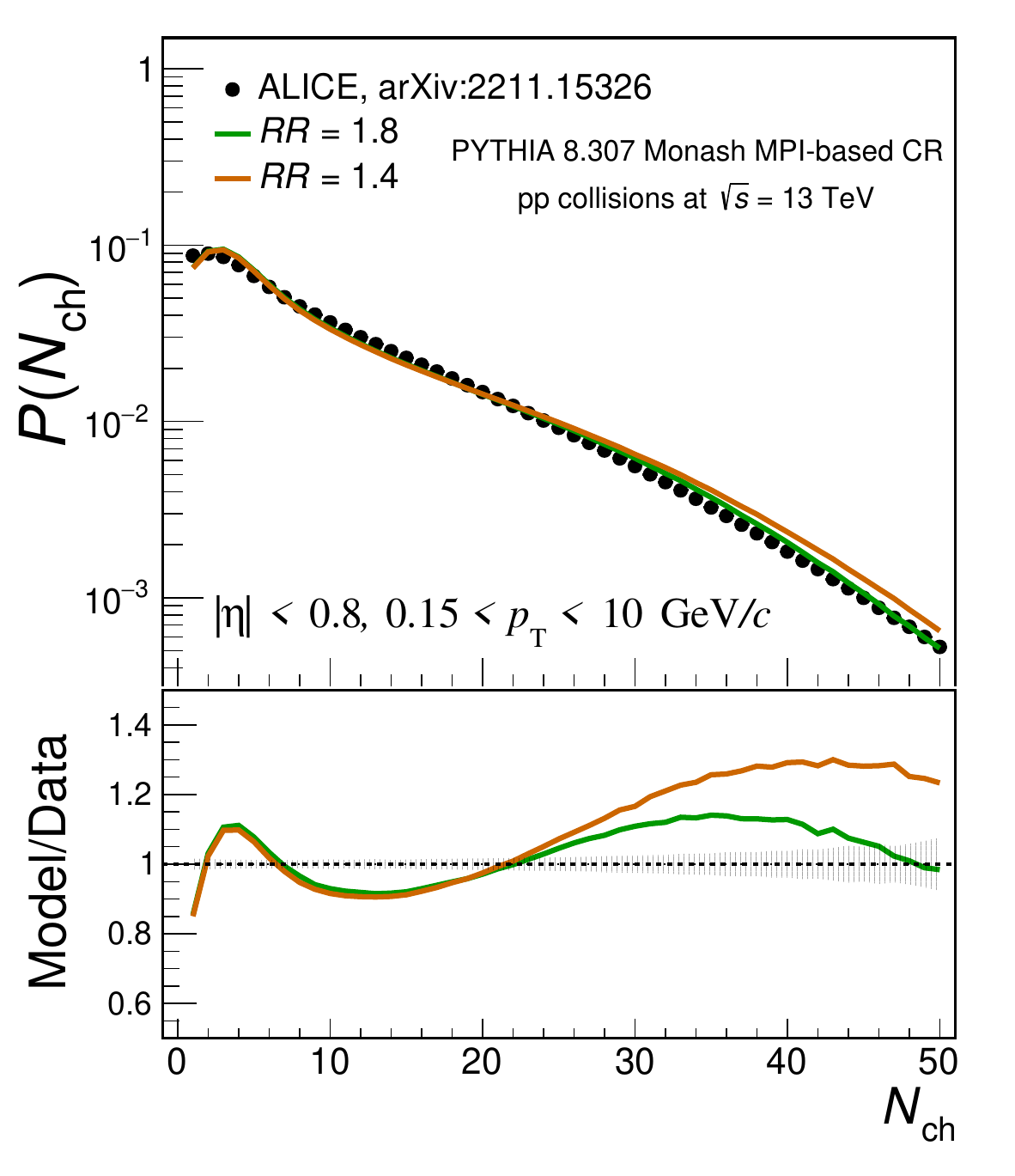}
\caption{Probability density of charged-particle multiplicity for \py Monash with $RR = 1.8$ (default) and $RR = 1.4$, denoted by the green and orange solid lines, respectively. Markers correspond to data and error bars are the associated systematic uncertainties. Bottom panel shows the model-to-data ratio.}
\label{Multiplicity}
\end{center}
\end{figure*}

Figure~\ref{Multiplicity} shows the probability density of charged-particle multiplicity for \py Monash with the reconnection rage values $RR=1.4$ and $RR=1.8$ (default). As mentioned above, a variation of $RR$ induces a modification of the particle production. As a matter of fact, both sets of simulations have the same results for d$N_{\mathrm{ch}}$/d$\eta$ $\lesssim$ 25. Above this value, setting $RR = 1.4$ over predicts the number of high-multiplicity pp collisions by a maximum of $\sim$ 20$\%$ relative to $RR = 1.8$ at ${\rm d}N_{\rm ch}/{\rm d}\eta=50$.

\subsection{Impact of jets on mean \pt}

Modification of  $RR$ improves the agreement between \py and data for jetty events, and it  affects the isotropic ones. Besides that, as already stated, the third increase in \meanpt at high-multiplicity (${\rm d}N_{\rm ch}/{\rm d}\eta>30$) is still observed even varying the $RR$ parameters. We therefore also studied the impact of jets in the discrepancy between \py Monash and data because jets are expected to modify the spectral shape at high \pt ($4<\pt<10$\,GeV/$c$). Bear in mind that CR effects on jet observables are expected to be negligible, because for example in the MPI-based CR model is easy to merge a low-$p_{\rm T}$ system with any other, but difficult to merge two high-$p_{\rm T}$ ones with each other~\cite{Bierlich:2022pfr}.

A recent publication has reported the inclusive charged-particle jet differential cross-section as a function of the jet-\pt for different jet radius ($R$) in pp collisions at $\sqrt{s}=13$\,TeV for two configurations: with and without background subtraction~\cite{JetMultiplicity}. For both set of results and small values of $R$, \py Monash overestimates the minimum-bias data up to  40$\%$ at $p_{\mathrm{T,jet}}$ $\approx$ 10\,GeV/$c$ and by 10\% at very high-$p_{\mathrm{T,jet}}$. The discrepancy is even larger if $R$ is increased. This is a clear indication that \py Monash overestimates the jet yield. Moreover, the discrepancy gets worst with increasing charged particle multiplicity. This could in turn explain the overpredicted \meanpt for large d$N_{\mathrm{ch}}$/d$\eta$ in jetty events. A procedure aimed at quantifying the effect of  such jet excess on \meanpt is followed.

\begin{figure*}
\begin{center}
\includegraphics[width=0.5\textwidth]{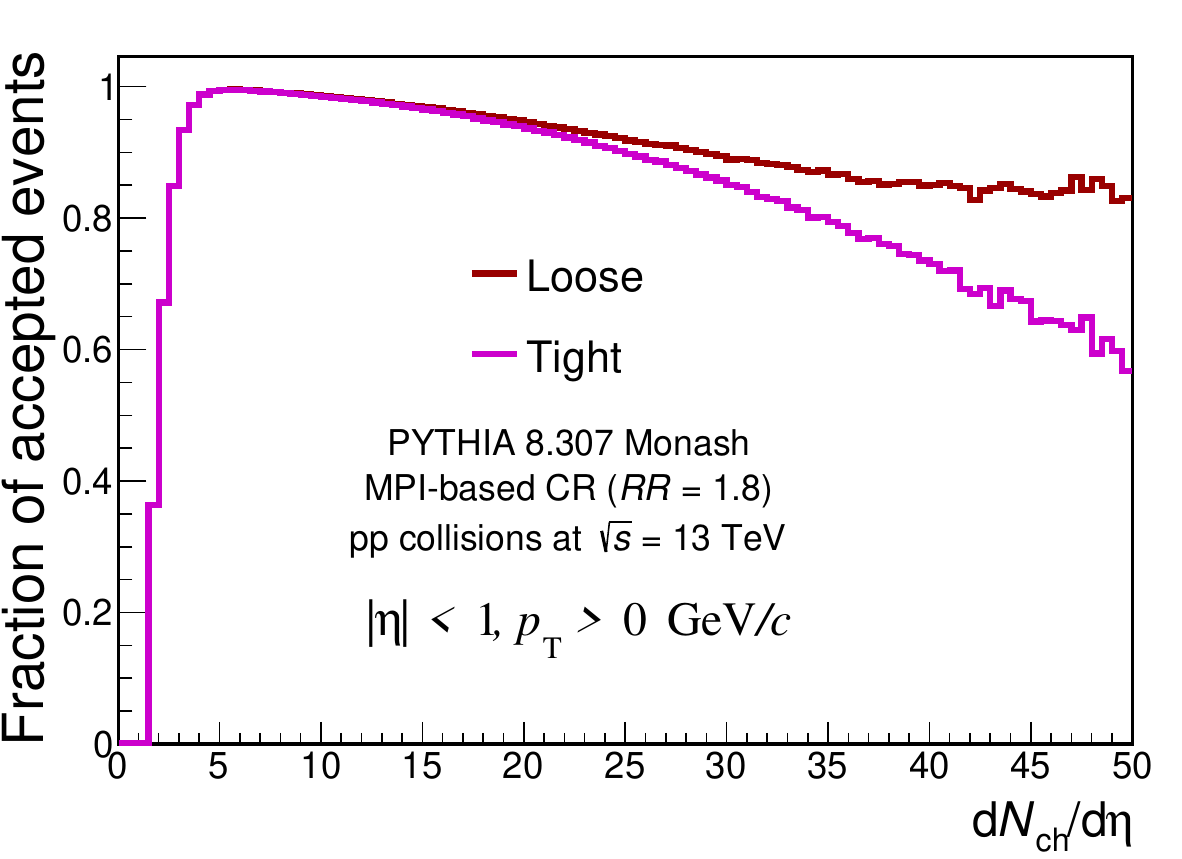}
\caption{Fraction of accepted events as a function of the charged-particle multiplicity. Accepted events are obtained applying the survival probability considering the loose (red) and tight (magenta) requirement.}
\label{AccRatio}
\end{center}
\end{figure*}

\py Monash simulations are employed using similar event and track selection described above with some modifications: no upper limit for particle's transverse momentum and  $|\eta|< 0.9$. FastJet is used as jetfinder with anti-$k_{\mathrm{T}}$ recombination algorithm \cite{FastJet}. Transverse momentum for jets is calculated with a boosted invariant \pt recombination scheme and $| \eta_{\mathrm{jet}}| < 0.9 - R$ (in this case $R=0.2$ is used). The jet-\pt spectrum measured by ALICE is normalized to the corresponding MC prediction. For $p_{\mathrm{T,jet}} >$ 5 \gevc, a survival probability given by this ratio is assigned to each jet-\pt interval. At this point two different selections are applied. The first one keeps events with at least one surviving jet (loose condition) and the second one requires events with all their jets surviving the selection (tight condition). Figure~\ref{AccRatio} shows the fraction of accepted events applying the survival probability to the leading jet (loose selection), and to all jets in the event (tight selection). Since low-multiplicity events are associated to soft physics like diffractive events, the hard selection removes most of the events for ${\rm d}N_{\rm ch}/{\rm d}\eta<5$. This is not an issue for our discussion because we are interested in the impact of jets at high multiplicities. The rejection factor is higher for the tight selection than for the loose selection at high multiplicity. 

\begin{figure*}
\begin{center}
\includegraphics[width=0.34\textwidth]{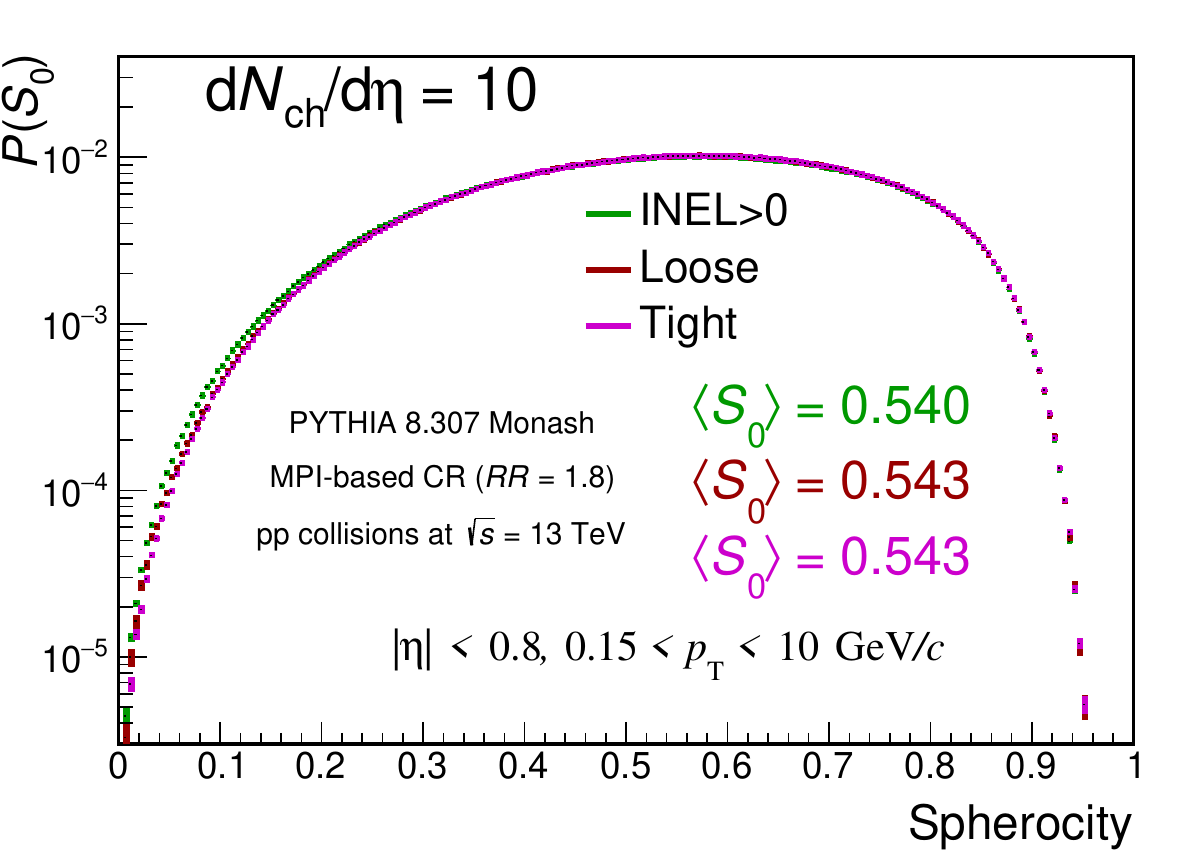}
\hspace{-0.4cm}
\includegraphics[width=0.34\textwidth]{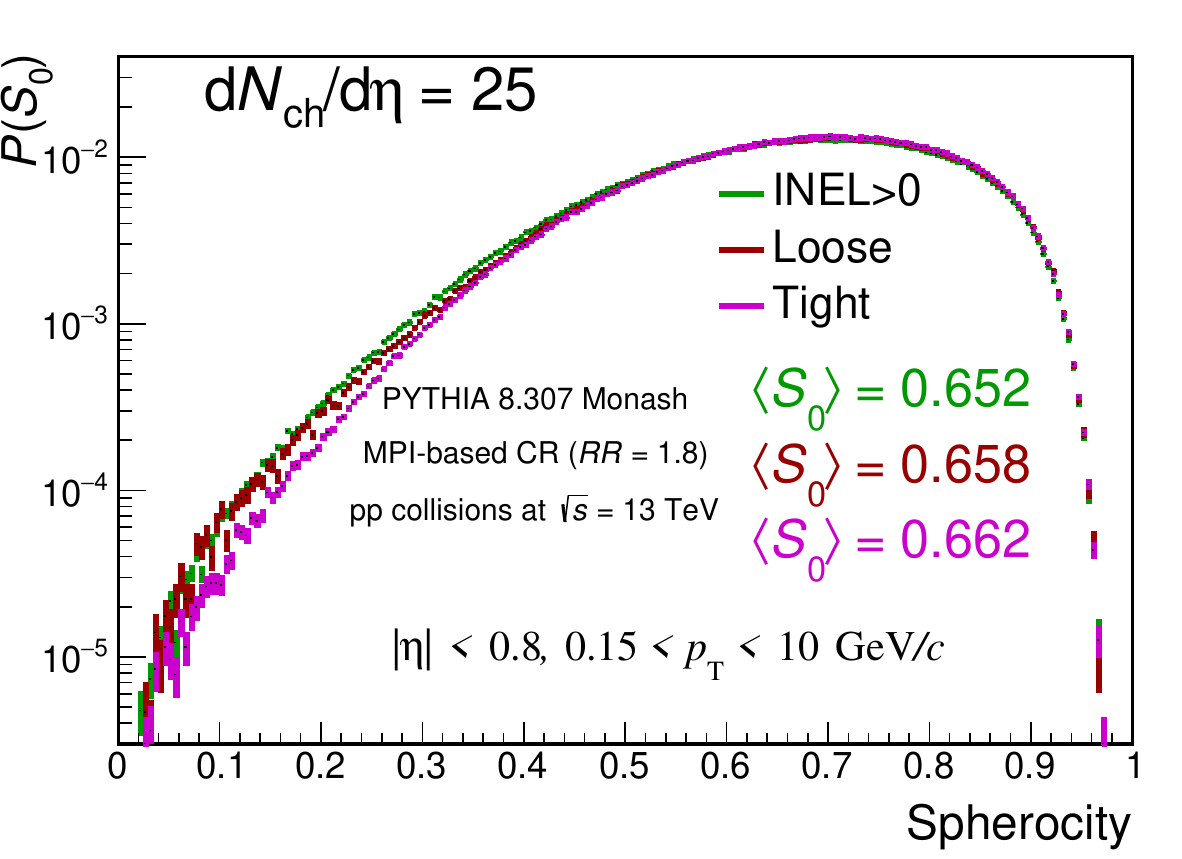}
\hspace{-0.4cm}
\includegraphics[width=0.34\textwidth]{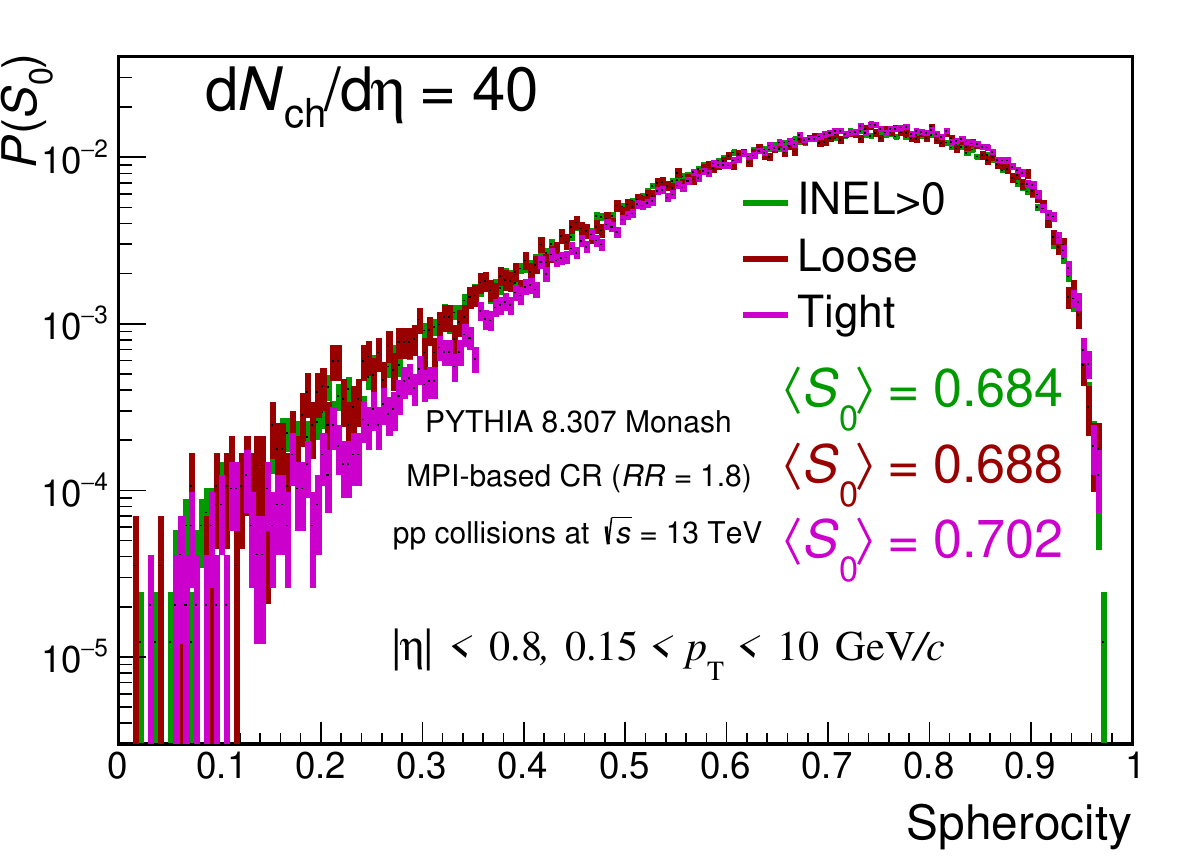}
\caption{\py Monash spherocity distributions for low (\textit{left}), intermediate (\textit{center}) and high (\textit{right}) multiplicity. Green line displays the spherocity-integrated result and the other two lines are obtained for events surviving the jet excess removal: red for the loose selection and magenta for the tight one. The corresponding mean values of the distributions are also shown.}
\label{S0Percentiles}
\end{center}
\end{figure*}

The selected events are in turn used to compute the \meanpt as a function of d$N_{\mathrm{ch}}$/d$\eta$ in the two spherocity classes. As the ``excess'' of jets has been removed, it is expected that jetty events become depleted in the \spher distributions shifting the average spherocity towards values closer to one. Figure~\ref{S0Percentiles} presents the spherocity distributions for low (\textit{left}), intermediate (\textit{center}) and high (\textit{right}) multiplicities considering INEL$>$0 and those surviving the loose and tight selections. Only results considering \py Monash are shown.  For quantitative comparison, the mean values for each \spher distributions are also displayed. These plots confirm the expected behaviour already stated: jetty events are now depleted towards larger \spher values. Furthermore, the effect is more pronounced as the multiplicity increases.

\begin{figure*}
\begin{center}
\includegraphics[width=0.49\textwidth]{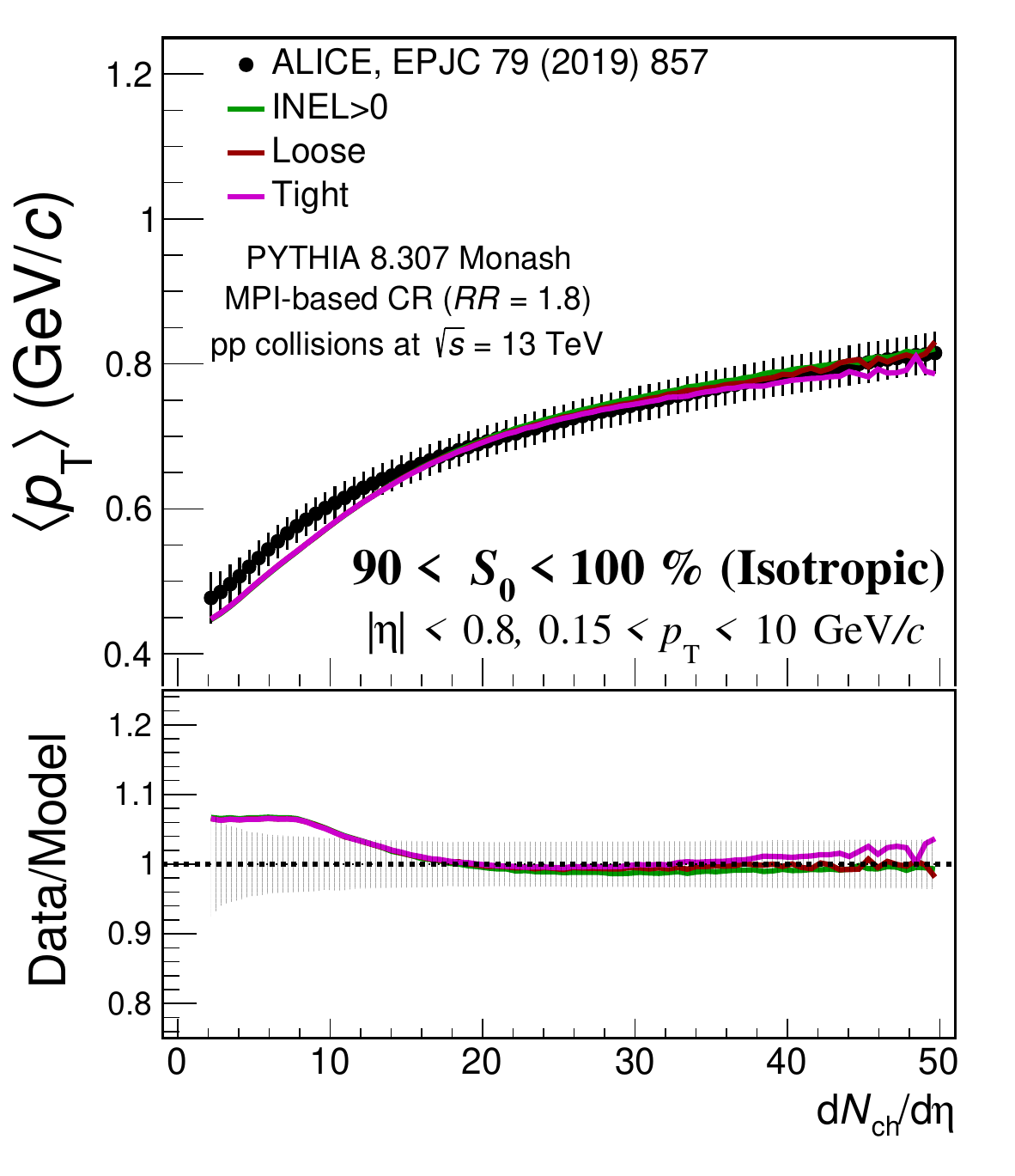}
\hspace{-0.4cm}
\includegraphics[width=0.49\textwidth]{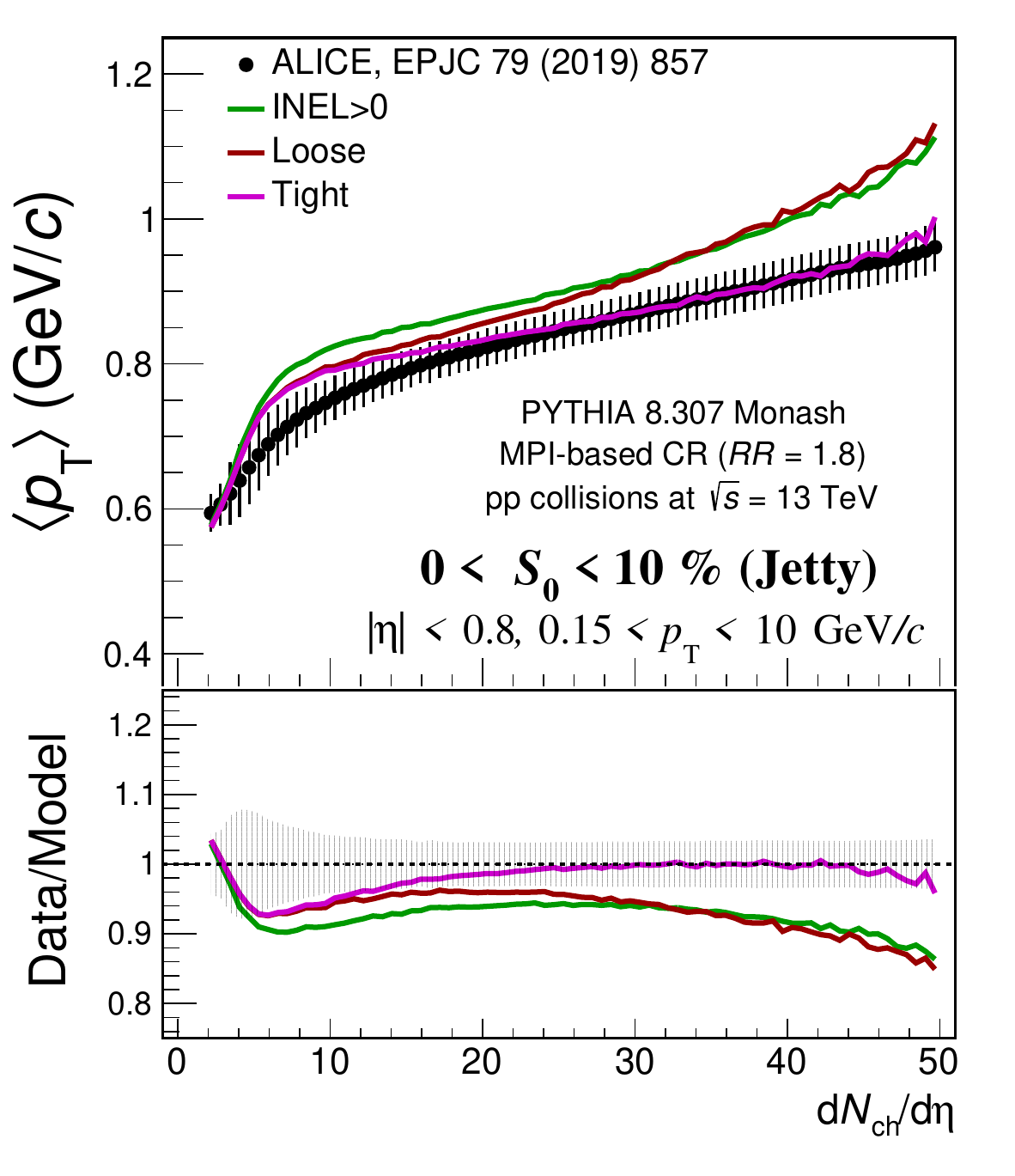}
\caption{Average \pt  as a function of d$N_{\mathrm{ch}}$/d$\eta$ for pp collisions at \s = 13 TeV for two different spherocity classes: isotropic (\textit{left}) and jetty (\textit{right}). Black markers correspond to data and error bars are the associated systematic uncertainties. Data are compared to three different Monte Carlo predictions (solid lines). Green line displays the result considering INEL$>$0 collisions and the other two lines are obtained for events surviving the jet excess removal: red for the loose selection and magenta for the tight one. Bottom panel presents the data to model ratio, where the shaded area around unity is the systematic uncertainty.}
\label{FilteredModels}
\end{center}
\end{figure*}

Figure \ref{FilteredModels} presents \meanpt as a function of d$N_{\mathrm{ch}}$/d$\eta$ in pp collisions at $\sqrt{s}=13$\,TeV for isotropic and jetty events. \py Monash results including INEL$>$0 collisions are compared with results obtained after applying the loose and tight selections. The prediction for isotropic events remains completely unaffected for both loose and tight selections, while for jetty events there is a noticeable difference. Regarding the loose selection, the result is closer to data as compared to INEL$>$0 for ${\rm d}N_{\rm chy}/{\rm d}\eta<30$, above this threshold both results are basically the same. Regarding the tight selection, the average \pt as a function of multiplicity is modified in the full multiplicity interval. Indeed, with this selection the model is able to  nicely describe data within systematic uncertainties.  These results suggest that for \py Monash to describe the data, the high-multiplicity regime should be dominated by minijet topologies rather than multijet final states. A similar conclusion was reached applying a Machine Learning technique to data, which suggested that in data multiparton interactions are more relevant than in MC to produce high multiplicities~\cite{Ortiz:2021peu}. In other words, the bias towards hard physics is not the same in data and MC.   

\section{Conclusion}

A study of the \meanpt as a function of charged particle multiplicity and spherocity has been presented. The main goal is to understand the origin of the discrepancy between data and \py Monash for pp collisions with spherocity close to zero (jetty events). There, the average \pt in \py Monash exhibits a steep increase with increasing multiplicity (${\rm d}N_{\rm ch}/{\rm d}\eta>30$) that is not seen in data. This effect is called third increase of the average \pt with multiplicity. This paper reports that the overestimate from \py Monash at high-charged-particle multiplicity is mainly due to to high-\pt particles (4 $<$ \pt $<$ 10 GeV/\textit{c}). Different color reconnection models were tested, as well as the impact of jets.  

\begin{itemize}
    \item Regarding color reconnection, we show that slightly decreasing the reconnection range parameter notably reduces the discrepancy, improving the agreement for jetty events but still keeping the third increase of the average \pt with multiplicity. The reduction affects the results for isotropic pp collisions in the full measured multiplicity range. Other predictions like multiplicity distributions are still in agreement with data within 20\%. Results from a new color reconnection model based on QCD rules yield same predictions as \py Monash for both jetty and isotropic events. 
    \item Regarding the impact of jets at high multiplicity, \py Monash is known to overestimate the jet production, in particular at high multiplicities. Based on comparisons between jet yields measured in INEL$>$0 collisions, the jet excess in \py Monash relative to data was estimated and used to get a rough estimate of the potential impact of this discrepancy on \meanpt. We define a survival probability of the event based on a data-motivated selection criterion applied to the leading jet. With this implementation, \py Monash keeps the very good description of the data in isotropic events but it reconciles the simulation with experimental measurements for d$N_{\mathrm{ch}}$/d$\eta$ $\lesssim$ 30 in jetty events. If the selectivity in \py Monash is applied to both leading and subleading jets, the agreement between data and \py Monash for jetty events gets significantly improved in the full multiplicity interval. The results suggest that the third rise of the average \pt for ${\rm d}N_{\rm ch}/{\rm d}\eta>30$ in \py Monash can be attributed to the presence of multijet topologies. The implication is that, in data, high multiplicities may be dominated by minijet topologies (MPI) rather than by multijet final states.  
\end{itemize}

\section{Acknowledgments}
Support for this work has been received from CONAHCyT under the Grants CB No. A1-S-22917, A1-S-21560 and CF No. 2042. 

\section*{References}

\bibliography{mybibfile}

\end{document}